\documentclass[12pt]{iopart}

\usepackage{bm}
\usepackage{epsfig}

\newcommand{\sigmaSI}{\sigma_{{\rm SI}}}

\newcommand{\kev}{{\rm keV}}

\newcommand{\gev}{{\rm GeV}}

\newcommand{\pb}{{\rm pb}}
\newcommand{\cm}{{\rm cm}}
\newcommand{\m}{{\rm m}}
\newcommand{\km}{{\rm km}}
\newcommand{\s}{{\rm s}}

\newcommand{\Gyr}{{\rm Gyr}}

\newcommand{\eqref}[1]{Eq.~(\ref{#1})}

\newcommand{\secref}[1]{Sec.~\ref{sec:#1}}

\newcommand{\figref}[1]{Fig.~\ref{fig:#1}}

\newcommand{\eg}{{\em e.g.}}
\newcommand{\gsim}{\lower.7ex\hbox{$\;\stackrel{\textstyle>}{\sim}\;$}}

\begin{document}



\title[Testing the Dark Matter Interpretation of DAMA/LIBRA with Super-K]
{Testing the Dark Matter Interpretation of the
DAMA/LIBRA Result with Super-Kamiokande}

\author{Jonathan L.~Feng}
\address{Department of Physics and Astronomy, University of
California, Irvine, CA 92697, USA
}

\author{Jason Kumar%
}
\address{Department of Physics and Astronomy, University of
California, Irvine, CA 92697, USA
}
\address{Department of Physics and Astronomy, University of
Hawai'i, Honolulu, HI 96822, USA
}

\author{John Learned}
\address{Department of Physics and Astronomy, University of
Hawai'i, Honolulu, HI 96822, USA
}

\author{Louis E. Strigari%
}
\address{Department of Physics and Astronomy, University of
California, Irvine, CA 92697, USA}
\address{Kavli Institute for Particle Astrophysics and Cosmology, 
Department of Physics, Stanford University, Stanford, CA 94305, USA} 


\begin{abstract}
We consider the prospects for testing the dark
matter interpretation of the DAMA/LIBRA signal with the
Super-Kamiokande experiment.  The DAMA/LIBRA signal favors dark matter
with low mass and high scattering cross section.  We show that these
characteristics imply that the scattering cross section that enters
the DAMA/LIBRA event rate determines the annihilation rate probed by
Super-Kamiokande.  Current limits from Super-Kamiokande through-going
events do not test the DAMA/LIBRA favored region.  We show, however,
that upcoming analyses including fully-contained events with
sensitivity to dark matter masses from 5 to 10 GeV may corroborate the
DAMA/LIBRA signal.  We conclude by considering three specific dark
matter candidates, neutralinos, WIMPless dark matter, and mirror dark
matter, which illustrate the various model-dependent assumptions
entering our analysis.
\end{abstract}


\maketitle

\section{Introduction}

The DAMA/LIBRA experiment has seen, with $8.2\sigma$
significance~\cite{Bernabei:2008yi}, an annual
modulation~\cite{Drukier:1986tm} in the rate of scattering events,
which could be consistent with dark matter-nucleon scattering.  Much
of the region of dark matter parameter space that is favored by DAMA
is excluded by null results from other direct detection experiments,
including CRESST~\cite{Angloher:2002in}, CDMS~\cite{Akerib:2005kh},
XENON10~\cite{Angle:2007uj}, TEXONO~\cite{Lin:2007ka,Avignone:2008xc},
and CoGeNT~\cite{Aalseth:2008rx}.  On the other hand, astrophysical
uncertainties~\cite{Brhlik:1999tt,Gondolo:2005hh} and detector
effects~\cite{Bernabei:2007hw} act to open up regions that may
simultaneously accommodate the results from DAMA and these other
experiments.  Following DAMA's latest results, several recent
studies~\cite{Feng:2008dz,Petriello:2008jj,Chang:2008xa,%
Fairbairn:2008gz,Savage:2008er} have studied the consistency of DAMA
with other direct detection experiments, with varying assumptions and
varying conclusions.  What is clear, however, is that if DAMA is
seeing dark matter, the preferred region of parameter space has dark
matter mass in the range $m_X \sim 1-10~\gev$ and spin-independent
proton scattering cross section $\sigmaSI \sim 10^{-5} - 10^{-2}~\pb$.
Although neutralinos have been proposed as a possible
explanation~\cite{Bottino:2003iu}, such low masses and high cross
sections are not typical of weakly-interacting massive particles
(WIMPs), and alternative dark matter candidates have been suggested to
explain the DAMA signal~\cite{Smith:2001hy,Feng:2008ya,Feng:2008mu,%
Foot:2008nw,Feng:2008dz,Khlopov:2008ty,Andreas:2008xy,Dudas:2008eq}.

The current state of affairs also makes it abundantly clear that data
from complementary experiments is likely required to sort out the true
nature of this result and is certainly required to establish
definitively the detection of dark matter.  Other direct detection
experiments may play this role.  In this paper, we note that
corroborating evidence may come from a very different source, namely,
from the indirect detection of dark matter at Super-Kamiokande
(Super-K).  In contrast to direct detection experiments, which rapidly
lose sensitivity at low masses, given physical limitations on
threshold energies, Super-K's limits remain strong for low masses.
Super-K is therefore poised as one of the most promising experiments
to either corroborate or exclude many dark matter interpretations of
the DAMA/LIBRA data.

In \secref{relating}, we show that, with a few well-motivated
theoretical assumptions, the DAMA and Super-K event rates may be
related.  Currently published Super-K results do not challenge the
DAMA preferred region.  In \secref{projection}, however, we show that
there is significant potential for Super-K to extend its reach to dark
matter masses from 5 to 20 GeV and provide sensitivity that is
competitive with, or possibly much better than, direct detection
experiments.  In \secref{models}, we apply our analysis to three
specific dark matter candidates that have been proposed to explain
DAMA: neutralinos~\cite{Bottino:2003iu}, WIMPless dark
matter~\cite{Feng:2008ya,Feng:2008dz,Feng:2008mu}, and mirror dark
matter~\cite{Foot:2008nw}.  These candidates illustrate and clarify
the assumptions entering the analysis.  We present our conclusions in
\secref{summary}.

As this work was in preparation, a study appeared that also considered
testing the dark matter interpretation of DAMA with data from
Super-K~\cite{Hooper:2008cf}. That work focused primarily on a
model-independent approach and present Super-K data, considering
neutralinos briefly as a case example, whereas this work considers
neutralino, WIMPless, and mirror dark matter and emphasizes the much
brighter prospects for future Super-K results.

\section{Relating Super-Kamiokande to DAMA}
\label{sec:relating}

Super-K may indirectly detect dark matter by finding evidence for dark
matter annihilation in the Sun or Earth's core to standard model (SM)
particles that then decay to neutrinos.  In the case of muon
neutrinos, the essential idea is to use the observed rate of
upward-going muon events at Super-K to place an upper bound on the
annihilation rate of dark matter in the Sun or the Earth's core.  In
the low-mass region of interest, the dominant contribution to neutrino
production via dark matter annihilation is from the
Sun~\cite{Gould:1987ir}, and we will focus on the Sun below, although
the Earth may also provide an interesting signal.

The total annihilation rate is
\begin{equation}
\Gamma = {1\over 2} C \tanh^2 [ (aC)^{\frac{1}{2}} \tau ] \ ,
\end{equation}
where $C$ is the capture rate, $\tau \simeq 4.5~\Gyr$ is the age of
the solar system, and $a= \langle \sigma v \rangle /(4\sqrt{2} V)$,
with $\sigma$ the total dark matter annihilation cross section and $V$
the effective volume of WIMPs in the Sun ($V = 5.7 \times
10^{30}\,{\rm cm}^3 (1\,{\rm GeV} /
m_X)^{3/2}$~\cite{Gould:1987ir,Gould:1987ju,Hooper:2008cf}).  It has
been shown~\cite{Griest:1986yu,Gould:1987ir,Kamionkowski:1994dp} that
if $\langle \sigma v \rangle \sim 10^{-26}~\cm^3~\s^{-1}$, as required
for the thermal relic density of dark matter to be in the observed
range, then for the range of parameters used in this work the Sun is
in equilibrium (with an equilibration time $\sim$ 420 million years),
and $\Gamma \approx {1\over 2}C$.  The thermalization of captured dark
matter in the core of the sun occurs much more
rapidly~\cite{Spergel:1984re,Jungman:1995df}.  For the dark matter
mass $\gsim 4\,{\rm GeV}$, which we assume here, WIMP evaporation is
not relevant~\cite{Griest:1986yu,Gould:1987ju,Hooper:2008cf}.

The dark matter capture rate is~\cite{Gould:1987ir}
\begin{equation}
C = \left[ \left( {8\over 3\pi} \right)^{1\over 2} \sigma
{\rho_X \over m_X} \bar v   {M_B \over m} \right]
\left[ {3\over 2} {\langle v^2 \rangle \over
\bar v^2} \right] f_2 f_3 \ .
\label{capturerate}
\end{equation}
The first bracketed factor counts the rate of dark matter-nucleus
interactions: $\sigma$ is the dark matter-nucleus scattering cross
section, $\rho_X/m_X$ is the local dark matter number density, $m$ is
the mass of the nucleus, and $M_B$ is the mass of the capturing
object.  The mean velocity of the dark matter is $\bar v$, and
$\langle v^2 \rangle$ is the squared escape velocity averaged
throughout the Sun.  The second bracketed expression is the
``focusing'' factor that accounts for the likelihood that a scattering
event will cause the dark matter particle to be captured.  The
parameters $f_2$ and $f_3$ are computable ${\cal O}(1)$ suppression
factors that account for the motion of the Sun and the mismatch
between $X$ and nucleus masses, respectively.  $f_3$ will be close to
1 for solar capture~\cite{Gould:1987ir}.  The point is that given
astrophysical assumptions about the density and velocity distribution
of dark matter, the capture rate is completely computable as a
function of the ratio $\sigma/m_X$.  Roughly, assuming $\rho_X =
0.3~\gev~\cm^{-3}$, $\bar v \sim 300{{\rm km} \over {\rm s}}$,
${3\over 2} {\langle v^2 \rangle \over \bar v^2} \sim 20
$~\cite{Gould:1987ir}, and taking $f_2 \sim f_3 \sim 1$, one finds $C
\sim 10^{29}~(\sigma/m_X)~\gev~\pb^{-1}~\s^{-1}$.

The major remaining particle physics uncertainty is the neutrino
spectrum that arises from dark matter annihilation.  This information
is encoded in the function~\cite{Jungman:1994jr}
\begin{equation}
\xi(m_X) =\sum_F
B_F [3.47 \langle N z^2 \rangle_{F,\nu} +2.08 \langle Nz^2
\rangle_{F,\bar \nu}] \ ,
\label{xi}
\end{equation}
which relates the dark matter annihilation rate to the rate of muon
events at a detector.  The $B_F$ are the branching fractions to each
of the $F$ SM final states summed over, and the $\langle N z^2
\rangle$ are the second moments of the (anti-)neutrino energy
spectrum, $dN/dE$, (normalized to the energy of the SM final state)
for the given dark matter mass and SM final
state~\cite{Kamionkowski:1994dp}.  $N$ is the total number of
neutrinos produced, derived from the neutrino energy
distribution. Note that, for the energies we consider, typical RMS
neutrino energies are $\sim 0.3 m_b$.  The scaling in eq.~\ref{xi} can
be understood by noting that both the muon range and the
neutrino-nucleon cross section are proportional to neutrino energy at
the energies we consider. Assuming the dark matter annihilates only to
SM particles, a conservative estimate for neutrino production may be
obtained by assuming that the annihilation to SM particles is
dominated by $b \bar b$ production for $m_b < m_X < M_W$, by $\tau
\bar \tau$ production for $m_W < m_X < m_t$, and by $W,Z$ production
for $m_X > m_t$~\cite{Jungman:1994jr}.

\begin{figure}
\resizebox{5.0in}{!}{
\includegraphics{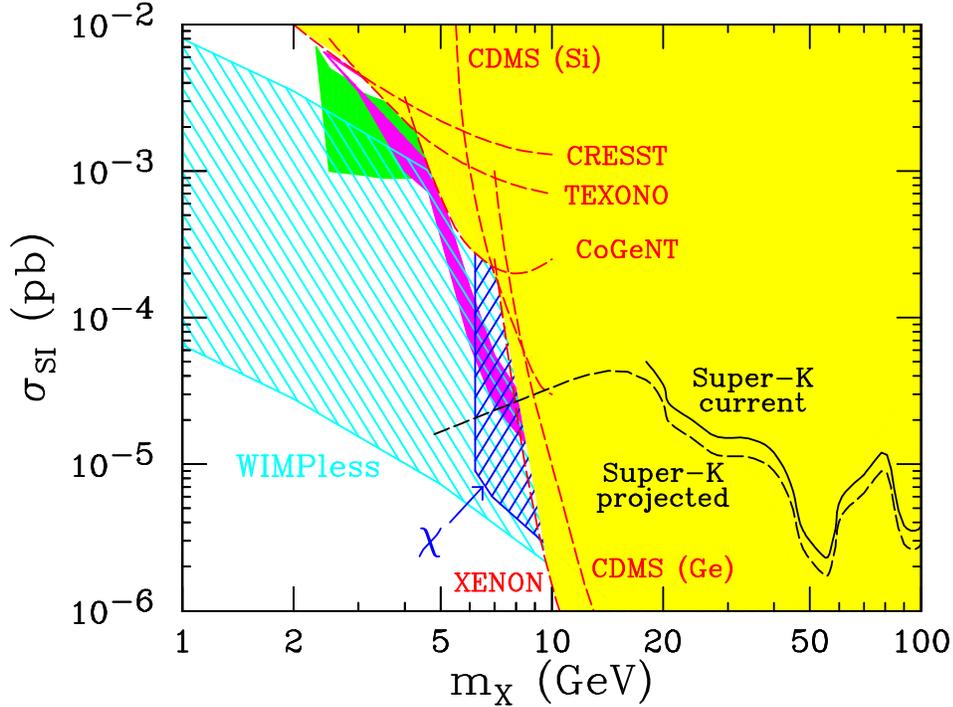}
}
\caption{Direct detection cross sections for spin-independent
$X$-nucleon scattering as a function of dark matter mass $m_X$.  The
black solid line is the published Super-K exclusion
limit~\cite{Desai:2004pq}, and the black dashed line is our projection
of future Super-K sensitivity. The magenta shaded region is
DAMA-favored given channeling and no streams~\cite{Petriello:2008jj},
and the medium green shaded region is DAMA-favored at 3$\sigma$ given
streams but no channeling~\cite{Gondolo:2005hh}.  The light yellow
shaded region is excluded by the direct detection experiments
indicated.  The dark blue diagonal shaded (upper right to lower left)
region is the prediction for the neutralino models considered in
Ref.~\cite{Bottino:2003iu} and the light blue diagonal shaded region
(upper left to lower right) region is the parameter space of WIMPless
models with connector quark mass $m_Y = 400~\gev$ and $0.3 <\lambda_b
< 1.0$.  Other limits come from the Baksan and MACRO experiments
\cite{Montaruli:1999nv,delosHeros:2007hy,Desai:2004pq}, though they
are not as sensitive as Super-K.
\label{fig:supdirect}
}
\end{figure}

Super-K bounds the $\nu_{\mu}$-flux from dark matter annihilation in
the Sun.  Since the total annihilation rate is equal to the capture
rate, this permits Super-K to bound the dark matter-nucleon scattering
cross section using \eqref{capturerate}, assuming $\rho_X =
0.3~\gev~\cm^{-3}$ and a Maxwellian velocity distribution with $\bar v
\sim 220~\km/\s$.  In \figref{supdirect}, we plot the published bounds
from Super-K, as well as limits from other dark matter direct
detection experiments and the regions of $(m_X, \sigmaSI)$ parameter
space favored by the DAMA signal given various astrophysical and
detector uncertainties.  As evident from \figref{supdirect}, the
published Super-K bounds (solid line) do not yet test the DAMA-favored
regions.  In the following section, however, we will see that
consideration of the full Super-K event sample may provide marked
improvements, and extend Super-K's sensitivity to low masses and the
DAMA regions.

\section{Projection of Super-K Sensitivity}
\label{sec:projection}

The Super-K inner detector has a radius of 16.9 meters and a height of
36.2 meters.  A muon event which is produced outside the inner
detector, passes through and then leaves the inner detector is called
a through-going muon.  These events are subject to a cut requiring the
muon to pass through at least 7 meters of the inner detector before
exiting.

As shown in \figref{supdirect}, Super-K currently reports dark matter
bounds only down to $m_X = 18~\gev$.  The reason is that more massive
dark matter particles produce energetic neutrinos that convert to
energetic muons.  For heavy dark matter, these muons are energetic
enough that they pass all the way through the detector, thus providing
maximal directional information, which can be used to distinguish
these neutrinos from those that arise from atmospheric background.
For dark matter above 18 GeV, the authors of Ref.~\cite{Desai:2004pq}
estimate that more than 90\% of the upward-going muons will be
through-going.  However, one could study dark matter at lower masses
by using stopping, partially contained, or fully contained muons, that
is, upward-going muons that stop within the detector, begin within the
detector, or both.

To determine rough projected bounds one may obtain from these event
topologies, we adopt the following strategy.  We begin by
conservatively assuming that the measured neutrino spectrum at low
energies matches the predicted atmospheric background.  In any given
bin with $N$ neutrino events, the bound on the number of neutrinos
from dark matter annihilation is then $\sqrt{N}$.  This implies a
bound on the dark matter annihilation rate to neutrinos, which, for a
conservative choice of $\xi(m_X)$, yields a bound on the capture rate
and therefore a bound on $\sigma / m_X$.  To include experimental
acceptances and efficiencies, we scale our results to the published
limits at $m_X = 18~\gev$, assuming these effects do not vary greatly
in extrapolating down to the $5-10~\gev$ range of interest.

The annihilation of dark matter particles $X$ produces neutrinos with
typical energies between 1/3 and 1/2 of $m_X$.  The Sun is effectively
a point source of neutrinos.  The directions of the observed muons,
however, lie in a cone around the direction to the Sun, with rms
half-angle of approximately\footnote{This approximation is a rough
estimate based on kinematics, but will be sufficient for our
purposes.} $\theta = 20^\circ
\sqrt{10~\gev/E_{\nu}}$~\cite{Jungman:1995df}.  In
Ref.~\cite{Desai:2004pq}, bounds on dark matter with $m_X = 18~\gev$
were set using neutrinos with energies $E_{\nu} \sim 6-9~\gev$.  The
event sample used consisted of 81 upward through-going muons within a
$22^{\circ}$ angle of the Sun collected from 1679 live days.

To extrapolate to the masses $m_X \sim 5-10~\gev$ of interest, we must
consider neutrinos with energies between 2-4 GeV.  {}From
Ref.~\cite{Ashie:2005ik}, one can see that at these energies, the
detected events are dominantly fully-contained events, and so we use
this event topology.  The expected number of fully-contained events
from all directions is given in Ref.~\cite{Ashie:2005ik} in bins with
width 1 GeV.  To determine the number of relevance to us, we restrict
these events to those within the required cone around the Sun.  The
number of events is then
\begin{equation}
N_{solar} = N \frac{1-\cos\theta}{2} \ ,
\end{equation}
where $N$ is the total number of fully-contained events expected in
the $2-4\,{\rm GeV}$ energy range and $\theta$ is the cone opening
angle appropriate for that range.  We find $N = 168$ fully contained
events per 1000 live days.

We then convert this limit on event rate to a limit on the neutrino
flux by dividing by the effective cross section for the Super-K
experiment in the relevant energy range.  The effective cross-section
for Super-K to neutrinos in a particular energy range is an exclusive
cross-section to a particular sample of muon events (for example,
throughgoing or fully contained) which are observed at Super-K.  It is
defined as the the rate of Super-K muon events within that sample and
energy range divided by the atmospheric flux of neutrinos over that
energy range.  The effective cross-section can be estimated from
Ref.~\cite{Ashie:2005ik} by dividing the estimated rate of events by
the predicted atmospheric flux, integrated over the relevant range of
energies.  For fully contained events with neutrino energies between 2
and 4 GeV, we find that the effective cross section\footnote{One can
verify this by computing the neutrino- nucleon scattering
cross-section and multiplying by the number of nucleons in the inner
detector.  The cross-section depends on the first moment of the
neutrino spectrum, $\langle Nz \rangle$, which can be estimated
from~\cite{Jungman:1994jr}.  This calculation confirms the estimate
from~\cite{Ashie:2005ik}. }  is $\sim 2.1 \times 10^{-8}~\m^2$.  For
upward through-going events with neutrino energies $\sim 8~\gev$, the
effective cross-section is $\sim 1.7 \times 10^{-8}~\m^2$.

Assuming that the neutrino events are detected primarily in either the
fully-contained ($2-4~\gev$) or through-going sample, we can then set
the following $2\sigma$ limits on the time-integrated neutrino flux
due to dark matter annihilation:
\begin{eqnarray}
\Phi_{{\rm FC}}^{{\rm max}} &=& {2 \sqrt{N_{{\rm FC}}}
\over 2.1 \times 10^{-8}~\m^2 } \sim 1.6 \times 10^{9}~\m^{-2} \,
\sqrt{\frac{N_{{\rm days}}}{1679}}
\nonumber\\
\Phi_{{\rm TG}}^{{\rm max}} &=& {2 \sqrt{N_{{\rm TG}}}
\over 1.7 \times 10^{-8}~\m^2 } \sim 1.0 \times 10^{9}~\m^{-2} \,
\sqrt{\frac{N_{{\rm days}}}{1679}} \ ,
\end{eqnarray}
where $N_{{\rm FC}} = 168\, (N_{{\rm days}} / 1000)$ and $N_{{\rm TG}}
= 81\, (N_{{\rm days}} / 1679)$ are the number of fully-contained and
through-going events within the angle and energy ranges, respectively,
scaled to $N_{{\rm days}}$ live days.

The ratio of these flux limits obtained from the fully-contained and
through-going samples are then equal to the ratio of $\sigma / m_X$ in
the $5-10~\gev$ regime to the same quantity at $18~\gev$.  We find
\begin{equation}
{1.6 \times 10^{9}~\m^{-2} \over 1.0 \times 10^{9}~\m^{-2}}
\sim
\left({\sigma_{5-10} \over m_{5-10}}\right) \left({\sigma_{18}
\over 18~\gev}\right)^{-1} \ ,
\end{equation}
where $\sigma_{5-10}$ is the Super-K bound on the dark matter nucleon
cross-section for a dark matter particle with mass in the range
$5-10~\gev$, and $\sigma_{18}$ is the bound for a dark matter particle
with mass $18~\gev$.  In \figref{supdirect} this projected Super-K
bound is plotted, assuming 3000 live days of the SK I-III run.

We can check this projection by computing the bound on the
annihilation rate to neutrinos, and comparing it to the capture rate.
Given the expression for $\Phi_{{\rm FC}}^{{\rm max}}$ evaluated over
3000 live days and the distance from the sun to the earth, one finds
on abound on the annihilation rate to neutrinos given by
$\Gamma_{XX\rightarrow \nu_{\mu} \nu_{\mu}} < 2.3 \times 10^{24}\,{\rm
events/ s}$.  Approximating $\Gamma_{XX\rightarrow \nu_{\mu}
\nu_{\mu}} \sim \Gamma_{total} = {1\over 2} C$, we then find ${\sigma
\over m_X} < 6 \times 10^{-6} {{\rm pb} \over {\rm GeV}}$.  The
resulting approximate bounds are within a factor of two of those
plotted in \figref{supdirect}.

Note that, unlike direct detection experiments, this bound does not
become much worse at lower energies.  Indeed, Super-K may beat other
experiments at these energies, and the bound improves significantly
with time, in contrast to direct detection experiments, where the
bounds at low mass are essentially limited by energy thresholds.
Moreover, it is important to note that any individual model will have
its own specific neutrino energy spectrum, and analysis with this
spectrum in mind will enhance the sensitivity of Super-K to that
model.

We expect that the projected sensitivities derived here are
conservative for several reasons.  First, events from the Earth's core
may be included.  In fact, there may be hydrogen in the Earth's
core~\cite{Hydrogen}, leading to an enhancement of event rates at the
low masses of interest here.  Second, for the various data sets and
employing a single joint fit, one can use the expected angular
distribution of muon and electron products of the solar and Earth
neutrinos from dark matter annihilations, instead of a simple cone
which includes all (and too much background). This can be done as a
function of energy for the contained events (stopping and
through-going muons being all binned together). In addition, using
off-source fake cones can provide background checks free of
Monte-Carlo systematic error concerns. Of course, if any hint of
excess exists, one can start to test for such things as
$\nu_{\mu}/{\nu_e}$ ratio and the particle to anti-particle ratio (via
stopped muon decays and again perhaps employing on-source to
off-source systematic canceling tests).  Note however that, since for
dark matter masses in the few GeV range the angular size of the
capture region in the Earth will be large, one may have to worry about
confusing any terrestrial annihilations with atmospheric neutrino
oscillations.

\section{Prospects for Various Dark Matter Candidates}
\label{sec:models}

We now consider specific examples of theoretical models that have been
proposed to explain the DAMA result.  We present these both to
highlight various theoretical assumptions that have been made in the
previous discussion, and to provide concrete examples of what future
super-K analyses may tell us.

We first consider neutralino dark matter in supersymmetric models.
Although neutralinos typically have larger masses and lower cross
sections than required to explain the DAMA signal, special choices of
supersymmetry parameters may yield values in the DAMA-favored region.
Such models have been discussed in Ref.~\cite{Bottino:2003iu}.  In
particular, gaugino masses are not unified in these models, so that
neutralinos with masses below 10 GeV are not in conflict with chargino
masses bounds.

The region of the $(m_X, \sigmaSI)$ plane spanned by the models of
Ref.~\cite{Bottino:2003iu} that do not violate known constraints is
given in \figref{supdirect}.  The range in $\sigmaSI$ results largely
from nuclear uncertainties.  We see that if Super-K's limits can be
extended to lower mass and our conservative projection improved as
discussed above, Super-K could find evidence for models in this class.
We note, however, that these models are required only to have relic
densities that do not overclose the Universe; many of them have $\rho
< 0.3~\gev~\cm^{-3}$.  For these models, Super-K's bound on the cross
section will be less sensitive than reported under the assumption
$\rho = 0.3~\gev~\cm^{-3}$ (see \eqref{capturerate}).

WIMPless dark matter provides an alternative explanation of the
DAMA/LIBRA signal~\cite{Feng:2008dz}.  These candidates are hidden
sector particles that naturally have the correct relic
density~\cite{Feng:2008ya}.  In these models, the dark matter particle
$X$ couples to SM quarks via exchange of a connector particle $Y$ that
is similar to a 4th generation quark.  The Lagrangian for this
interaction is
\begin{equation}
{\cal L} = \lambda_f X \bar{Y}_L f_L
+ \lambda_f X \bar{Y}_R f_R \ .
\label{connector}
\end{equation}
The Yukawa couplings $\lambda_f$ are model-dependent, and it is
assumed that only the coupling to 3rd generation quarks is
significant, while the others are Cabbibo-suppressed.\footnote{This is
a reasonable assumption and is consistent with small observed
flavor-changing neutral currents.}  In this case, one finds that the
dominant nuclear coupling of WIMPless dark matter is to gluons via a
loop of $b$-quarks ($t$-quark loops are suppressed by $m_t$).  The
$X$-nucleus cross section is then given by~\cite{Feng:2008dz}
\begin{equation}
\sigmaSI = \frac{1}{4\pi}
\frac{m_N^2}{(m_N + m_X)^2}
\left[ \sum_q \frac{\lambda_b^2}{m_Y - m_X}
\left[ Z B^p_b + (A-Z) B^n_b \right] \right]^2 \ ,
\end{equation}
where $Z$ and $A$ are the atomic number and mass of the target nucleus
$N$, and $B^{p,n}_b = (2/27) m_{p} f^{p,n}_g / m_b$, where $f^{p,n}_g
\simeq 0.8$~\cite{Cheng:1988im,Ellis:2001hv}.

In \figref{supdirect}, the range in the $(m_X, \sigmaSI)$ parameter
space for WIMPless models with $m_Y = 400~\gev$ and $0.3 <\lambda_b <
1.0$ is also given.  We see that these models span a large range in
the $(m_X, \sigmaSI)$ plane, and may overlap all parts of the
DAMA-favored region.  In this case, since dark matter annihilation to
SM particles proceeds only through the $b\bar b$ channel, the
conservative Super-K estimate for $\xi(m_X)$, defined in \eqref{xi},
is largely correct.  We see that Super-K's projected sensitivity may
be sufficient to discover a signal that corroborates DAMA's.

WIMPless models illustrate an important caveat to the analysis above,
however.  In WIMPless models, dark matter may also annihilate to
hidden sector particles, which, of course, do not produce neutrinos
detectable at Super-K.  If there are hidden decay channels, then the
annihilation rate to SM particles is
\begin{equation}
\Gamma_{SM} = B(X\bar X \to {\rm SM}) \Gamma_{{\rm tot}} =
B(X\bar X \to {\rm SM})\, C \ ,
\end{equation}
and one should divide the Super-K limit by $B(X\bar X \to {\rm SM})$
to obtain the Super-K bound in the presence of hidden decay channels.

The cross section for annihilation to hidden sector particles cannot
be arbitrarily large, however, if the thermal relic density is to
remain significant.  In this WIMPless model, for $\lambda_b = 0.5$,
the cross-section for annihilation to SM particles is already $(\sigma
v)_{SM} \sim 7~\pb$.  WIMPless dark matter also annihilates to hidden
sector particles through hidden gauge interactions.  In contrast to
neutralinos and other visible sector dark matter candidates, however,
for WIMPless dark matter the precise relation between this
annihilation cross section and the relic density depends on some
model-dependent factors, such as the number of relativistic degrees of
freedom in the hidden sector and the ratio of hidden and visible
sector temperatures~\cite{Feng:2008mu}.  For reasonable values of
these parameters, the models plotted in \figref{supdirect} can have a
relic density that is $10-100\%$ of the observed density of dark
matter.  For the lower densities in this range, Super-K's sensitivity
will be proportionally worse, according to \eqref{capturerate}.  In
any case, if Super-K's reach can extend to lower mass, it could
reasonably find evidence for (or place constraints on) these models.

Finally, we consider mirror dark matter, which has also been advanced
as a DAMA-explaining possibility~\cite{Foot:2008nw}.  These models
require a slightly more detailed analysis.  In this case, the dark
matter candidate is a hidden sector particle with mass $\sim
10-30~\gev$ that interacts with the SM through kinetic mixing with the
photon.  Because scattering occurs through exchange of a massless
particle, one finds here that $\sigma \propto E_R ^{-2}$, where $E_R$
is the recoil energy.  This is why mirror dark matter can be seen at
DAMA while evading bounds from other direct detection experiments ---
the cross section is highest for scattering at low recoil energies,
which are below the threshold of other other experiments, but above
the 2 keV threshold of DAMA.

To understand the limits Super-K can place on this type of model, we
must essentially find the ``threshold'' of the Sun, thought of as a
direct detection device.  The Sun only captures particles that lose an
energy $E_0$ during elastic scattering, where $E_0$ is the kinetic
energy of the particle when it was far from the Sun.  $E_0$ is
essentially the threshold recoil energy required for dark matter
capture.  For a halo velocity $\sim 220~\km/\s$ and a mass $\sim
20~\gev$, we find $E_0 \sim 5~\kev$.  The solar capture rate is
unaffected (to first approximation) by scattering at lower recoil
energies; essentially, for the purposes of a mirror matter study,
Super-K is a high-threshold experiment\footnote{Note that this is in
contrast to Super-K probes of WIMPs, which Super-K can study at low
mass.  This difference arises because, with mirror matter, the issue
is the scaling of the cross-section with recoil energy.  The
suppression of the cross-section at high recoil energy does not
typically occur with WIMP models.}.  Indeed, the threshold is higher
than that of XENON10 (4.5 keV).  In the relevant mass range, a mirror
matter candidate that could explain the DAMA signal cannot be detected
by XENON10~\cite{Foot:2008nw}.  Since XENON10 has higher sensitivity
than Super-K in this mass region, and the cross-section relevant for
Super-K is slightly smaller than that for XENON10, one expects that
Super-K also will not be able to bound mirror matter.  We note,
however, that models with $m \sim 10~\gev$ may be constrained by
Super-K.

\section{Summary}
\label{sec:summary}

The DAMA/LIBRA signal is currently of great interest, and alternative
methods for corroborating or excluding a dark matter interpretation
are desired.  In this study, we have shown that the preferred DAMA
region implies that Super-K, through its search for dark matter
annihilation to neutrinos, has promising prospects for testing DAMA.

We have given a conservative estimate of the projected sensitivity of
Super-K.  By using fully contained events, we expect that current
super-K bounds may be extended to dark matter masses of 5 to 10 GeV.
In the region of most interest for the DAMA result with $m_X \sim
5-10~\gev$, the neutralino models of Ref.~\cite{Bottino:2003iu} and
WIMPless models can potentially be tested, provided the sensitivities
expected at this low mass range are actually realized.  For mirror
matter, however, the mass range of interest is $10-30~\gev$, and it is
unlikely that Super-K can place limits on this model.

We thus have the intriguing prospect that the direct detection result
of DAMA/LIBRA could be sharply tested by an indirect detection
experiment in the very near future.

\section*{Acknowledgments}

We are grateful to Huitzu Tu, Hai-Bo Yu and, especially, Hank Sobel
for discussions.  This work was supported by NSF grants PHY--0239817,
PHY--0314712, PHY--0551164 and PHY--0653656, DOE grant
DE-FG02-04ER41291, and the Alfred P.~Sloan Foundation.  Support for
this work for LES was provided by NASA through Hubble Fellowship grant
HF-01225.01 awarded by the Space Telescope Science Institute, which is
operated by the Association of Universities for Research in Astronomy,
Inc., for NASA, under contract NAS 5-26555. JK is grateful to CERN and
the organizers of Strings '08, where part of this work was done, for
their hospitality.

\vspace*{.5in}



\end{document}